\documentclass[journal,transmag]{IEEEtran}
\usepackage{graphicx}
\usepackage{mathptmx} 
\usepackage{amsmath} 
\usepackage{graphicx}
\usepackage{subfigure} 

\usepackage{color}
\ifCLASSINFOpdf
\else
\fi
\begin{document}
%
\title{Powertrain Hybridization for Autonomous Vehicles  \textsc{}}


\author{\IEEEauthorblockN{Shima Nazari, Norma Gowans, and
Mohammad Abtahi}
\IEEEauthorblockA{Department of Mechanical and Aerospace Engineering, UC Davis, Davis, CA 95618 USA}}


\IEEEtitleabstractindextext{%
\begin{abstract}
The powertrains of today's hybrid electric vehicles (HEVs) are developed for human drivers and, therefore, may not be the optimum choice for future Autonomous vehicles (AVs), given that AVs can accurately manipulate their velocity profile to avoid unnecessary energy loss.    In this work, we closely examine the necessary degree of hybridization for AVs compared to human drivers by deploying real-world urban driving profiles and generating equivalent AV drive cycles in a mixed autonomy scenario. We solve the optimal energy management problem for HEVs with various motor sizes from the automotive market, and demonstrate that while human drivers typically require a motor size of around 30 kW to fully benefit from hybridization, AVs can achieve similar gains with only a 12 kW motor. This greater benefit from a smaller motor size can be attributed to a more optimal torque request, allowing for higher gains from regenerative braking and a more efficient engine operation.
Furthermore, We investigate the benefits of velocity smoothing for both traditional cars and HEVs and explore the role of different mechanisms contributing to fuel consumption reduction. Our analysis reveals that velocity smoothing provides greater benefits to HEVs equipped with small motors compared to non-hybrid vehicles and HEVs with larger motors.

\end{abstract}

}

\maketitle

\IEEEdisplaynontitleabstractindextext

\IEEEpeerreviewmaketitle

\section{Introduction}


\IEEEPARstart{V}{ehicle}  electrification is seen as the most promising pathway towards reducing the carbon emissions of transportation sector. While battery technology continues to evolve, Hybrid Electric Vehicles (HEVs) are a suitable short-term and mid-term solution that integrate the benefits of traditional Internal Combustion Engine Vehicles (ICEVs) with Electric Vehicles (EVs). Hybridization effectively reduces a vehicle's fuel consumption through several mechanisms:  operating the engine at a higher efficiency, recovering the vehicle's kinetic energy through regenerative braking instead of relying  on friction brakes, and shutting down the engine when idling. Meanwhile, over the past decade a significant amount of research has been dedicated to vehicle automation. Automated Vehicles (AVs) offer numerous benefits to the society, such as improved safety, mobility, and comfort \cite{pettigrew2018health}. AVs enable a better planning for transportation resources and hence can reduce transportation costs \cite{litman2020autonomous}. Moreover, removing the driver from the vehicle control loop allows an enhanced management of the vehicle to encode efficient driving styles known as eco-driving \cite{barth2009energy, vagg2013development}. Eco-driving involves manipulating the vehicle's velocity profile for a lower fuel consumption \cite{huang2018eco}. While these driving practices can be followed by human drivers as well, \cite{andrieu2012comparing}, the implementation is more likely to be  effective when integrated with the longitudinal motion controller of an AV or in an Advanced Driver Assistance System (ADAS) such as an Adaptive Cruise Control (ACC) \cite{li2008mpc, klunder2009impact}. 

The effectiveness of hybridization in fuel consumption reduction is substantiated in the literature and in practice \cite{Srdjan2004, nazari2020power}, and the capability of vehicle velocity optimization and eco-driving for reducing energy consumption has been studied in the past for traditional ICEVs \cite{kato2013comparative}, HEVs, \cite{maamria2018computation,franke2016ecodriving}, and EVs \cite{dib2014optimal, zhang2015eco}. 
For example, Kato et al. \cite{kato2013comparative} showed that eco-driving can effectively improve the fuel economy of all mentioned types of vehicles through chassis dynamometer experiments. In another work, Manzie et al. \cite{manzie2007fuel} compared the fuel economy gains of vehicle hybridization versus velocity smoothing, achieved by traffic information preview over standard urban drive cycles. They found that the fuel economy improvement of a traditional ICEV from velocity smoothing with 60 seconds of preview is comparable to full vehicle hybridization. This result is not surprising because powertrain hybridization and velocity optimization work in similar ways to improve fuel economy; for example, they both avoid energy loss in friction brakes by regenerative braking (in HEVs) or less/smoother braking (for eco-driving). 

Given this analogy in concept, the fuel consumption reduction from these two approaches is not accumulative, yet, the synergy between powertrain hybridization and velocity optimization, achieved by automation and ADAS, is not fully understood. This work takes a closer look at this connection by comparing the fuel economy of a traditional ICEV and HEVs with human drivers, to the fuel economy of AVs with the same powertrains but with optimized velocity profiles. An initial study \cite{nazari2019effectiveness} showed that low-voltage 48 V systems have the potential for significant fuel consumption reduction if coupled with eco-driving over standard EPA cycles. This work expands the prior study by including real-world urban driving profiles and by considering the existing levels of hybridization in the automotive market. We retrofitted a Ford Escape powertrain with different electric motors in a P2 parallel hybrid structure to generate comparable HEV powertrains for this study. We selected the parallel architecture because it is the simplest and cheapest type of HEV, and existing ICEVs can easily be retrofitted to a parallel HEV by coupling a motor with the engine crankshaft. Note that the motor sizing problem for parallel HEVs with human drivers is studied in the past \cite{Srdjan2004,Freyermuth2009}. Nevertheless, this work, for the first time, investigates the necessary degree of hybridization for automated vehicles in a mixed autonomy scenario. 

\begin{figure}[htbp]
\centerline{\includegraphics[width=0.49\textwidth]{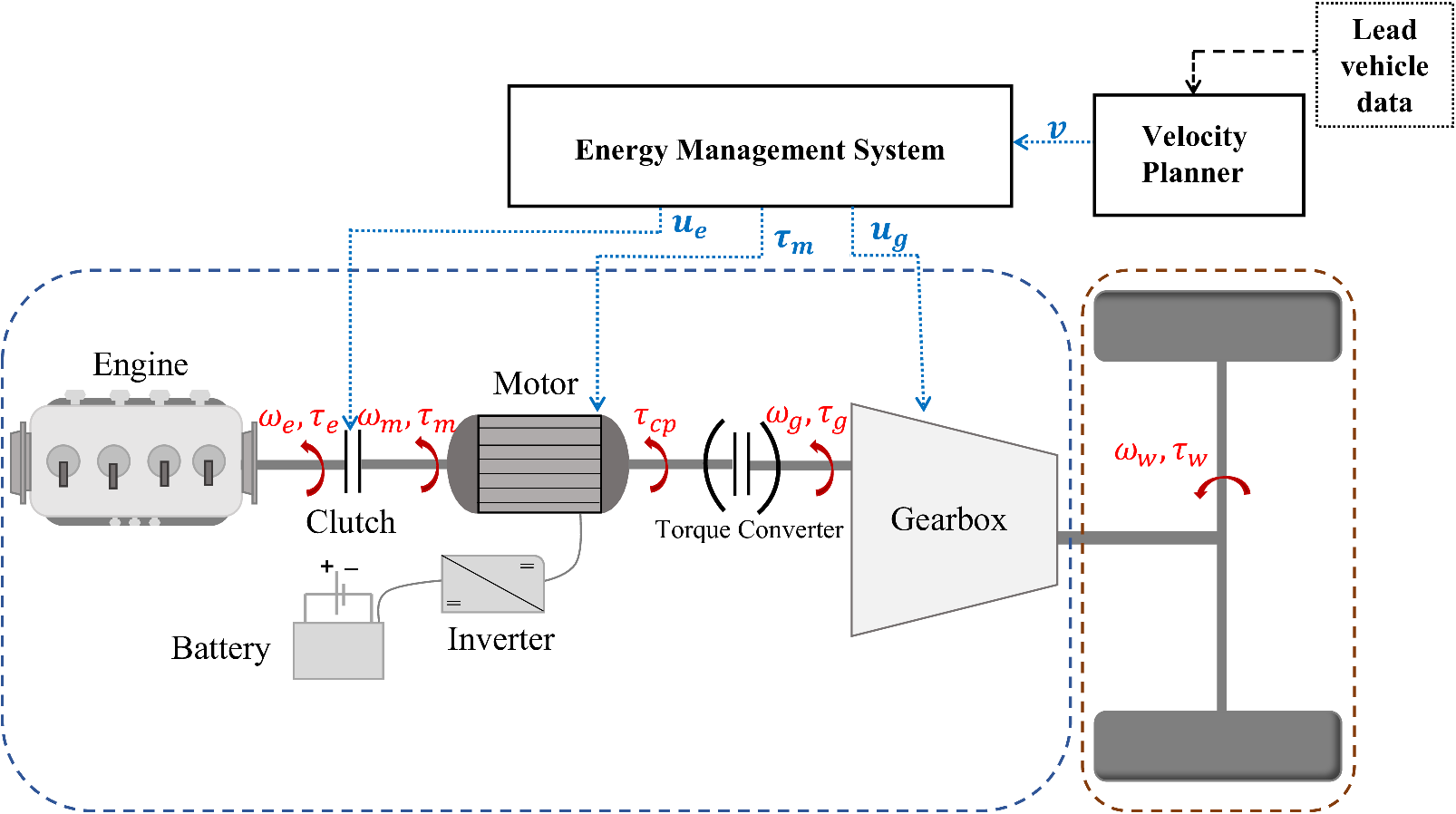}}
\caption{Schematic view of the parallel hybrid vehicle powertrain.}
\label{fig_powertrain}
\end{figure}

The true optimum fuel economy for HEVs is achieved by co-optimization of the power split ratio and vehicle acceleration \cite{heppeler2017predictive}. However, the co-optimization problem suffers from the curse of dimensionality and thus is computationally expensive, such that iterative approaches are usually adopted to find the global optimum solution \cite{chen2020iterative, padilla2018global, chen2022co}. In this work, we adopted a sequential optimization method in which the vehicle velocity profile in a presumed autonomous driving scenario is optimized first, and in the next step, the optimal energy management for the powertrain is solved as shown in Fig. \ref{fig_powertrain}. This approach, although sub-optimal, is computationally tractable and shown to be an effective and practical approach \cite{chen2018sequential}, especially for analyses with a large number of studied drive cycles such as here. We used Dynamic Programming (DP) to solve both the velocity profile optimization and the energy management problem. The optimal fuel economy numbers and powertrain parameters for different degrees of hybridization with Human Driver (HD) and AV are compared against each other to understand the relation between hybridization, vehicle automation, and fuel economy.

The reminder of this paper is organized as follows, first the vehicle and the studied powertrains are introduced in Section \ref{sec:powertrain}, followed by the real-world driving profiles  in Section \ref{sec:drive_profile}. Section \ref{sec:drive_cycle_AV} formulates the optimization problem for building the AV drive cycles in a car-following scenario, and Section \ref{sec:energy_management} defines the energy management problem. The vehicle and powertrain models used in this part are given in the Appendix. The results are discussed in Section \ref{sec:discussion} and the paper ends with concluding remarks.

\section{Studied Vehicle and Powertrains}
\label{sec:powertrain}
The baseline vehicle is a Ford Escape MY2015 with a 1.6L downsized turbocharged engine and a 6-speed automatic transmission. The data on the vehicle parameters and fuel consumption map is published elsewhere \cite{nazari2018assessing}. In this work we retrofitted this powertrain with a clutch and an electric motor in a P2 parallel hybrid structure, in which the motor is mounted between the engine and the gearbox and the clutch can decouple the engine from the rest of the drivetrain. A schematic of the resulting hybrid powertrain is shown in Fig. \ref{fig_powertrain}. Note that the boosted downsized engine is a suitable choice for this study because its efficiency sweet spot is wide and lies within the medium torque range, rather than the high torque area in non-boosted engines. Therefore,  over-sizing of the hybrid powertrains, due to using the same engine for all studied cases, will not be a  significant issue as shown later in this work. 

\begin{figure}[th!]%
\centering
\subfigure[][]{%
\label{motor_8}%
\includegraphics[trim= 0cm 0cm 0cm 0cm, clip=true, width=0.23\textwidth]{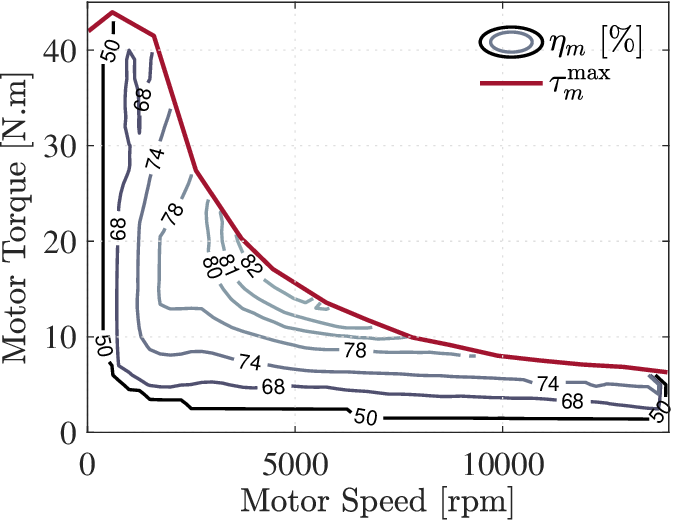} }
\subfigure[][]{%
\label{motor_12}%
\includegraphics[trim= 0cm 0cm 0cm 0cm, clip=true, width=0.23\textwidth]{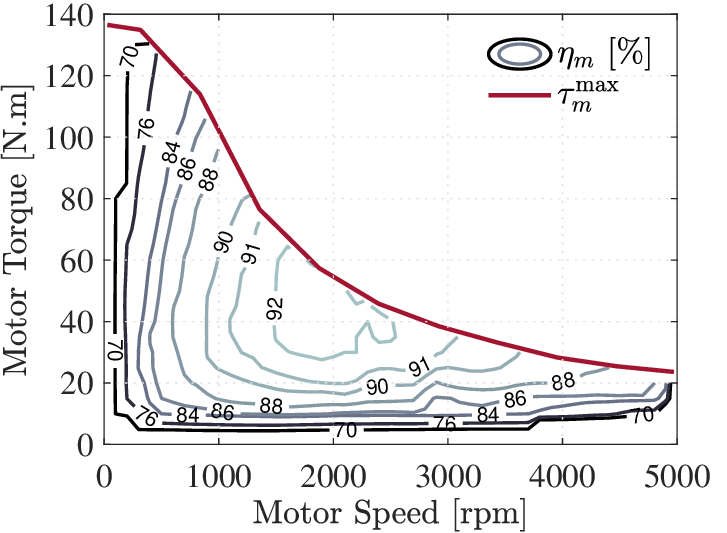} }
\subfigure[][]{%
\label{motor_30}%
\includegraphics[trim= 0cm 0cm 0cm 0cm, clip=true, width=0.23\textwidth]{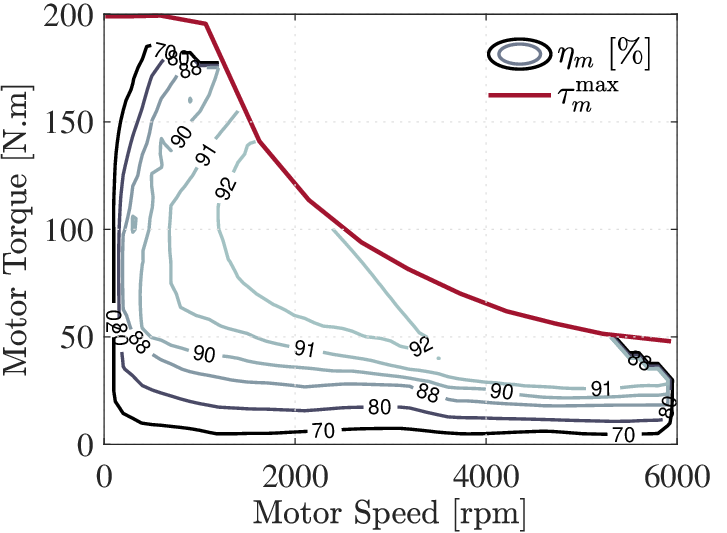} }
\subfigure[][]{%
\label{motor_40}%
\includegraphics[trim= 0cm 0cm 0cm 0cm, clip=true,  width=0.23\textwidth]{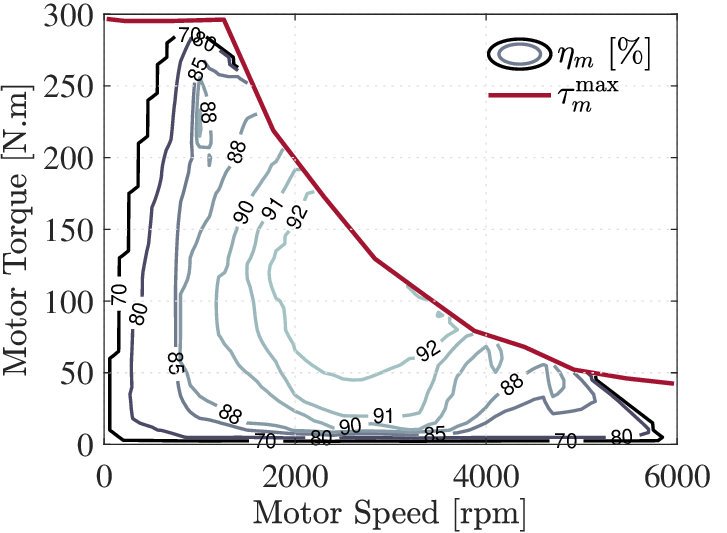} }
\caption[]{Combined motor-inverter efficiency map and peak torque curve for
\subref{motor_8} 8 kW motor from \cite{SonataMap1}, 
\subref{motor_12} 12 kW motor from \cite{staunton2006evaluation},
\subref{motor_30} 30 kW motor from \cite{SonataMap},
\subref{motor_40} 40 kW motor from \cite{burress2011evaluation}. }
\label{motor_maps}%
\end{figure}

Four electric motors, corresponding to various degrees of hybridization and found in published literature, were selected for this study. The 8 kW motor is from the Hyundai Sonata MY2012 starter-generator \cite{SonataMap1}, the 12 kW motor is from the Honda Accord MY2005 \cite{staunton2006evaluation},  the 30 kW motor is from the Hyundai Sonata MY2011 \cite{SonataMap}, and the last motor is from the Toyota Prius MY2010 \cite{burress2011evaluation}. It is stated that this motor produces 60 kW of power, however, the available peak torque curve used in this work generates around 40 kW maximum and therefore, we will refer to this motor as the 40 kW motor. Table \ref{Tab:mtr-tech} provides more information on the motors. The most relevant technical specifications are the peak torque curve and the motor-inverter combined electro-mechanical efficiency, both of which are given in Fig. \ref{motor_maps}. Two additional motor sizes are considered in this study. A 5 kW motor is realized by scaling down the 8 kW motor map and a 20 kW motor is built by scaling up the 12 kW motor map. In total, 6 different hybrid powertrains were considered in this study. A 1.2 kWh Li-ion battery is assumed for all hybrid vehicles.  





\begin{table}[!t]
	\caption[]{Technical details for the considered electric motors.}
	\centering
	\resizebox{0.45\textwidth}{!}{\begin{tabular}{|l| c| c| c| c|}
	 \hline
	Motor & Type & Peak Power [kW]  & Peak Motor-Inverter Efficiency \\
 
 \hline
 		  1& IPM & 5  & 82\% \\
		  2& IPM & 8  & 82\% \\
    	  3& PMSM & 12 & 92\% \\
          4& PMSM & 20 & 92\% \\
		  5& PMSM & 30  & 92\% \\
		  6& PMSM & 40  & 92\% \\
    \hline
		\end{tabular}}
	\label{Tab:mtr-tech}
\end{table}

\begin{figure}[h!]%
\centering
\subfigure[][Velocity]{%
\label{fig:vel_stat}%
\includegraphics[trim= 0cm 0cm 0cm 0cm, clip=true,  width=0.35\textwidth]{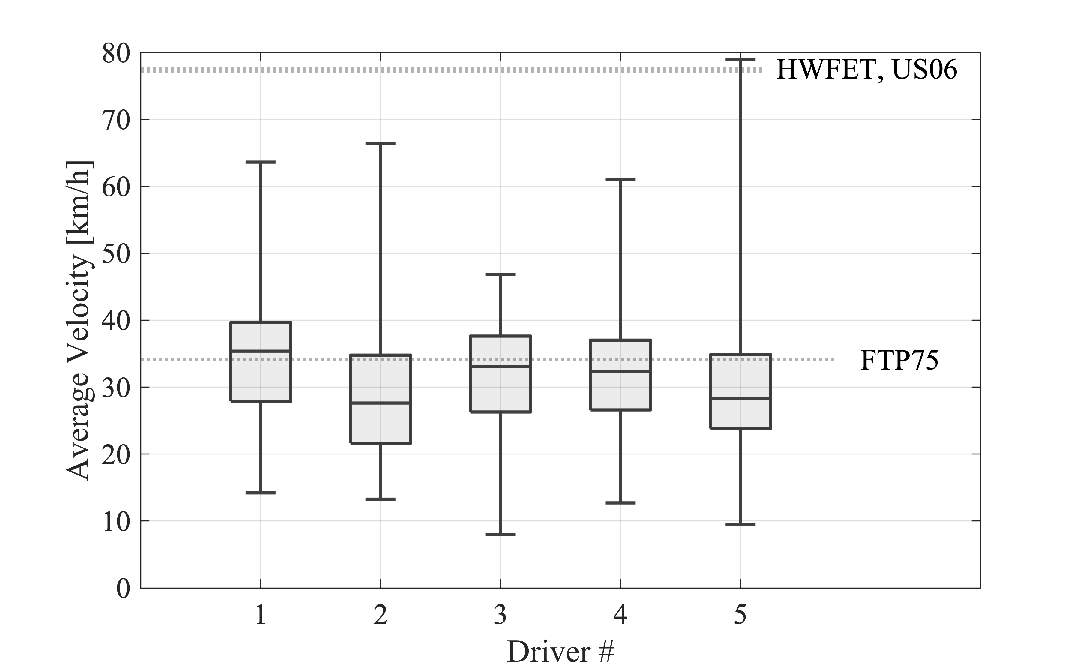} }
\subfigure[][Trip duration]{%
\label{fig:length_stat}%
\includegraphics[trim= 0cm 0cm 0cm 0cm, clip=true,  width=0.35\textwidth]{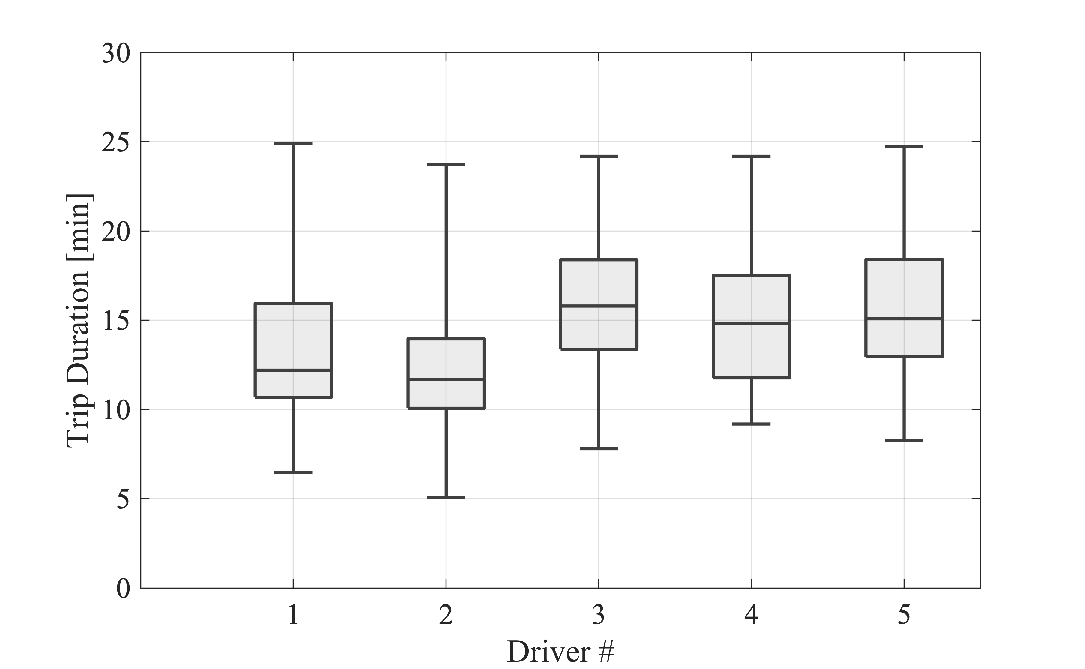} }
\caption[]{Statistics of the considered real-world driving profiles, the line in the center of the box shows the median value and the edges of the box show 25\% and 75\% of data for
\subref{fig:vel_stat} velocity, 
\subref{fig:length_stat} trip duration. }
\label{fig:profiles_stat}%
\end{figure}
\section{Real-world driving profiles}
\label{sec:drive_profile}
\begin{figure*}[tbh!]%
\centering
\subfigure[][acceleration]{%
\label{fig:d_5}%
\includegraphics[trim= 0cm 0cm 0cm 0cm, clip=true, width=0.35\textwidth]{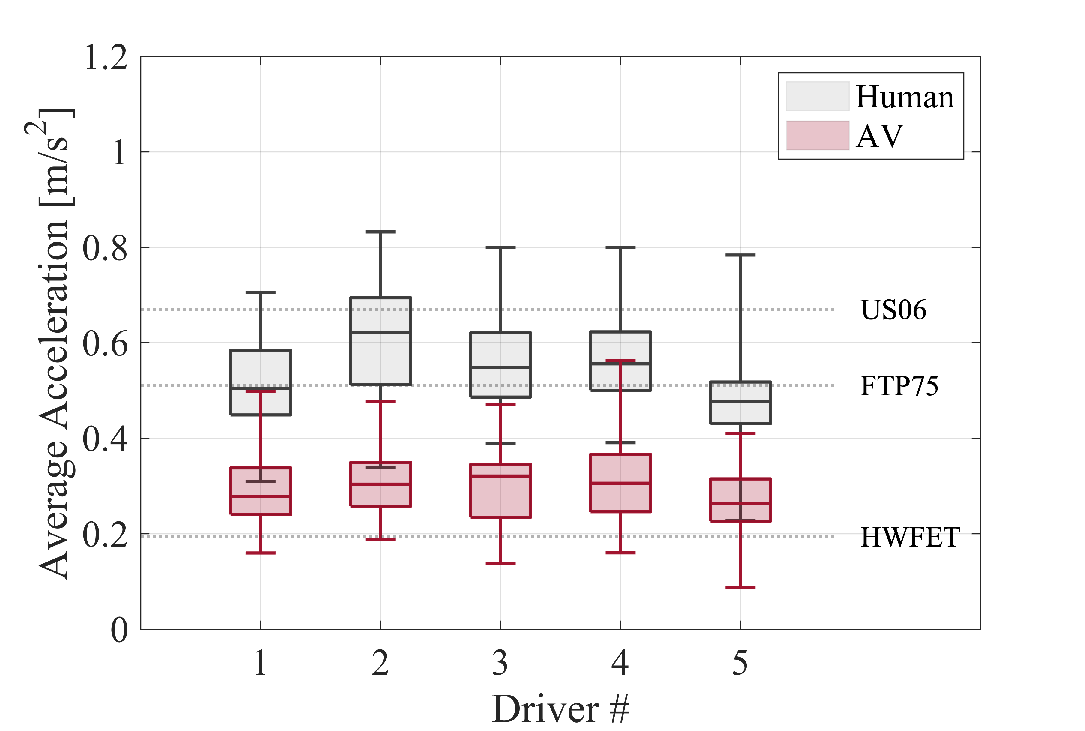} }
\subfigure[][deceleration]{%
\label{fig:d_3}%
\includegraphics[trim= 0cm 0cm 0cm 0cm, clip=true,  width=0.35\textwidth]{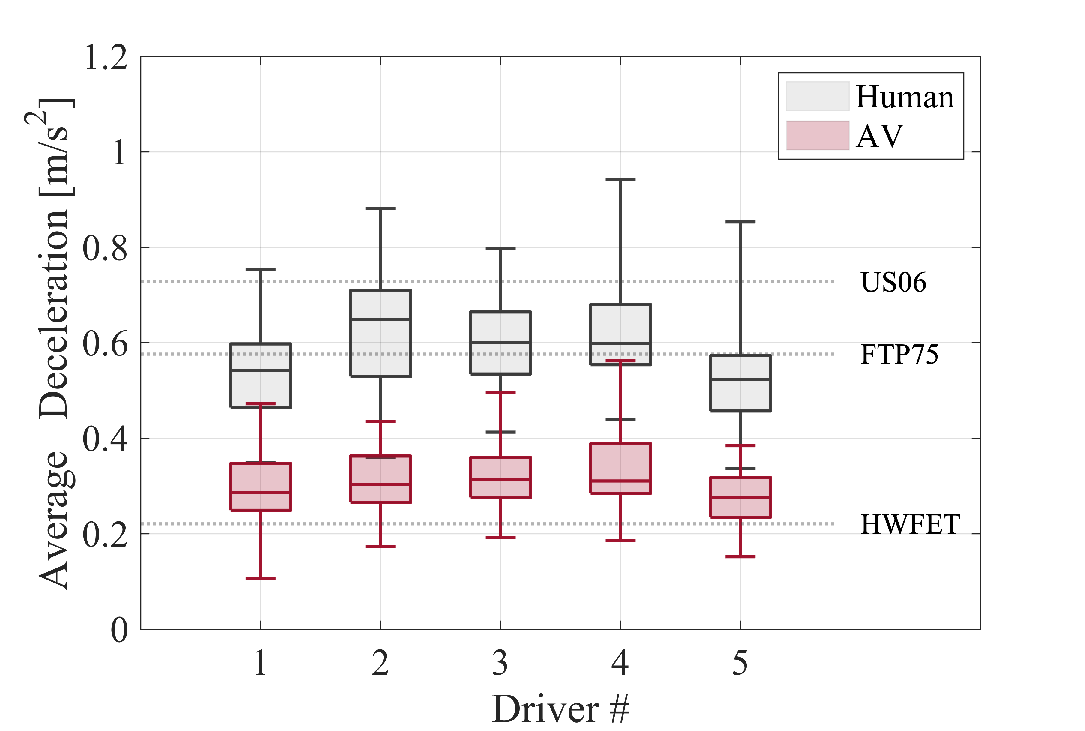} }
\caption[]{Real-world driving profiles  (grey) and optimized AV cycles (red),
\subref{fig:d_5} average acceleration distributions, 
\subref{fig:d_3} average absolute deceleration distributions }
\label{fig_acc_dist}%
\end{figure*}

\begin{figure}[t!]%
\centering
\subfigure[][]{%
\label{fig:DC_example1}%
\includegraphics[trim= 0cm 0cm 0cm 0cm, clip=true,  width=0.44\textwidth]{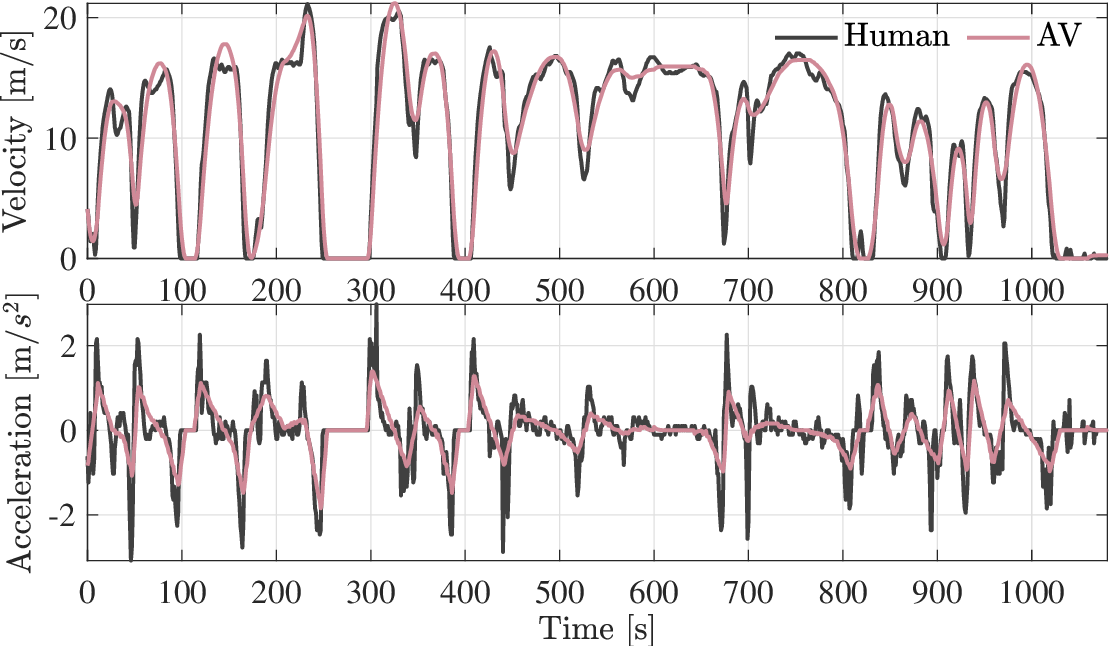} }\\
\subfigure[][]{%
\label{fig:DC_example2}%
\includegraphics[trim= 0cm 0cm 0cm 0cm, clip=true,  width=0.44\textwidth]{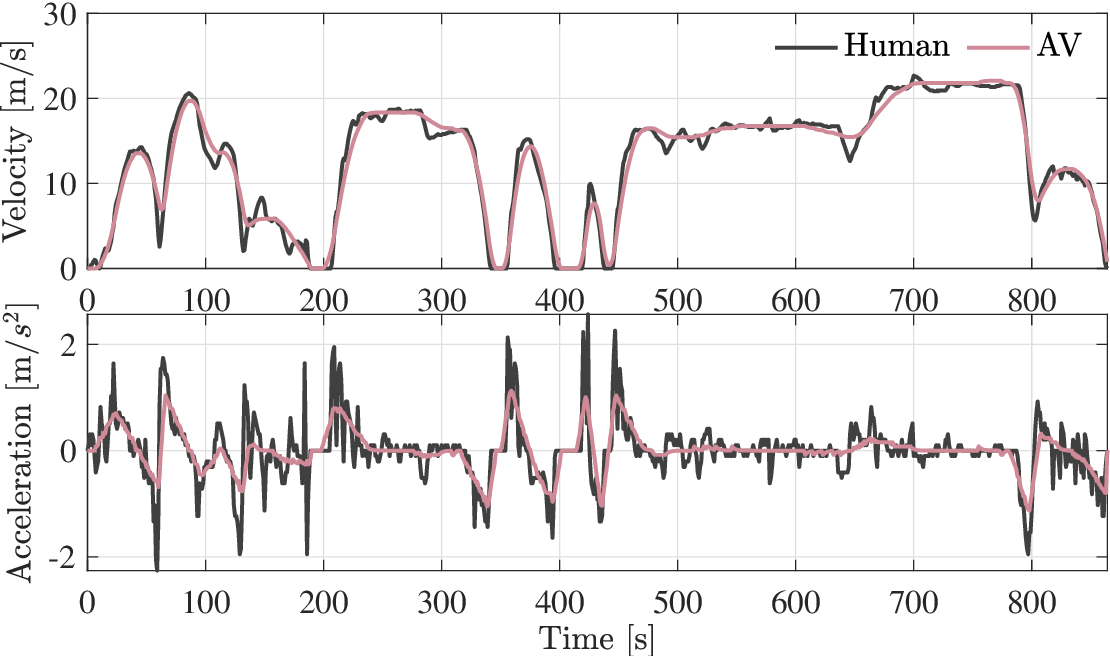} }
\caption[]{Examples of studied read-world cycles and modified cycles for autonomous car following,
\subref{fig:DC_example1} a cycle from driver 3, 
\subref{fig:DC_example2} a cycle from driver 2. }
\label{FE_Eco_trends}%
\end{figure}

In order to study the benefits of vehicle hybridization and automation and the synergy between these two in realistic conditions, this study adopted real-world urban driving profiles available for public use by Smith and Blair \cite{SmithDataSet, smith2011characterization}. The database provided the  dates, vehicle latitude and longitude coordinates, and vehicle speed with a 1-second time resolution. 270 cycles from five drivers were chosen for this study. Fig. \ref{fig:profiles_stat} provides  statistics on the cycle length and speed for each driver. The average velocity from EPA standard cycles, FTP75, HWFET, and US06, are also shown. As seen most cycles have an average velocity of less than 40 [$\rm km/h$] and the trip duration is less than 20 minutes. Fig. \ref{fig:d_5} and \ref{fig:d_3} show the statistics on the cycle acceleration and deceleration  for each driver as a measure of driving aggressiveness. The corresponding values for standard EPA  cycles  are also shown. Driver 2 has the largest average acceleration and absolute deceleration, and Driver 5 has the lowest values.  As seen, most drivers in terms of average velocity and aggressiveness are close to the FTP75 cycle. Examples of the speed profile for these cycles are given in Fig. \ref{fig:DC_example1}, \ref{fig:DC_example2}, and \ref{fig:HD_Cycle_EMS}.

\section{Drive cycles for autonomous cars}
\label{sec:drive_cycle_AV}
To compare the fuel economy of human-driven cars with AVs, an equivalent autonomous vehicle driving profile must be built for each human-driven cycle.  We adopted the method introduced by Prakash et. al \cite{prakash2016use}. This method first finds the velocity profile of a hypothetical lead vehicle for a  human-driven speed profile by inverting the Intelligent Driver Model (IDM) \cite{treiber2000congested}. In the next step, an optimization problem is solved to find the velocity profile of an autonomous vehicle that is following the same hypothetical lead vehicle. This approach minimizes the second norm of  acceleration  while keeping a safe distance from the lead vehicle that prohibits other vehicles' cut-in. This problem is formulated as,

\begin{align}
\label{eq:eco-drive}
\min \sum_1^N ||a(k)||_2^2
\end{align} 
subject to 
\begin{subequations}
\label{eq:vel_opt_limit}
		\begin{align}
		\label{eq:pos_dynamic}
		x(k+1) &= x(k) + v(k)T_s + \frac{1}{2}a(k)T_{s}^{2}\\
		\label{eq:vel_dynamic}
		v(k+1) &= v(k) + a(k)T_{s}\\
		\label{eq:acc_lim}
		a^{\rm min} &\leq a(k) \leq a^{\rm max}\\
		\label{eq:v_lim}
		v^{\rm min} &\leq v(k) \leq v^{\rm max}\\
		\label{eq:x_lim}
		x(k)^{\rm min} &\leq x(k) \leq x(k)^{\rm max}		
		\end{align}
\end{subequations}
Equations (\ref{eq:pos_dynamic}) and (\ref{eq:vel_dynamic}) portray the vehicle longitudinal kinematics, which is a second-order Linear Time-Invariant (LTI) model with acceleration, $a$, as system input and position, $x$, and velocity, $v$, as system states. The problem sampling time, $T_s$, is selected  equal to 1 second. The velocity and acceleration of the AV are limited by Equations (\ref{eq:acc_lim}) and (\ref{eq:v_lim}), and the distance from the hypothetical lead vehicle is regulated by the constraint (\ref{eq:x_lim}) at every moment to satisfy the above mentioned criteria. For more information on this method please see \cite{prakash2016use}. A MATLAB-based Dynamic Programming (DP) solver \cite{sundstrom2009generic} was used to find the equivalent autonomous vehicle profile for each of the considered cycles. Figures \ref{fig:d_5} to \ref{fig:d_3} compare the statistics of the AV cycles to human-driven cycles. As expected, the AV cycles have lower average acceleration and absolute deceleration, pointing to the smoother velocity profiles followed by autonomous vehicles. 
Fig. \ref{fig:DC_example1} and \ref{fig:DC_example2} compare two of the real-world cycles to their AV-driven profiles. As seen the velocity profiles are only slightly modified for a smoother following.

\section{Optimal energy management}
\label{sec:energy_management}
The Energy Management System (EMS) of an HEV must choose the share of the ICE and the electric motor in providing the requested torque by the driver/motion controller. Rule-based and optimization-based controllers are two main categories of EMS \cite{panday2014review}. Rule-based methods are simple and can be implemented in real-time, but they do not utilize knowledge of the drive cycle or a traffic preview, and hence, they produce sub-optimal results.  
Optimization-based approaches use physics-based models to optimize a performance metric, which is usually fuel consumption. Equivalent Consumption Minimization Strategy (ECMS) \cite{paganelli2002equivalent}, Model Predictive Control (MPC) \cite{borhan2011mpc} and DP \cite{lin2004stochastic} are among the most widely used optimization approaches for energy management of HEVs. Among these approaches, DP is usually used with a full drive cycle preview to produce a globally optimum solution. Such DP-based EMS is not practical for real-world implementation due to the high computational cost and lack of access to the entire vehicle velocity profile. However, it is a suitable approach to compare the relative effectiveness of different technologies in reducing the fuel consumption of a vehicle, such as in this study.

Backward simulation is employed in this study to model the HEV powertrain and driveline, with the detailed information provided in the Appendix. In this method, the velocity profile is an input to the model, and the wheels torque and rotational speed, $\tau_w$ and $\omega_w$ respectively (see Fig. \ref{fig_powertrain}), are computed using the given velocity and  vehicle parameters. These values are then back propagated along the driveline through the gearbox and torque converter to compute the torque and rotational speed on the motor shaft and the engine shaft. Eventually, the battery current, $I_b$, and State of Charge (SoC), $\zeta$, are computed using the motor torque, $\tau_m$, motor speed, $\omega_m$, and the motor efficiency. An Open Circuit Voltage model with a single Resistor (OCV-R) is used to model the battery. 
The problem inputs are the motor torque, gear shift command, $u_g$, and engine on/off command, $u_e$, and the modeled states are the battery state of charge, the gear number, $n_g$, and engine on/off state $x_e$. The fuel consumption minimization problem is formulated as,
\begin{figure*}[th!]%
\centering
\subfigure[][Human driven cycle]{%
\label{fig:HD_Cycle_EMS}
\includegraphics[trim= 0cm 0cm 0cm 0cm, clip=true, width=0.4\textwidth]{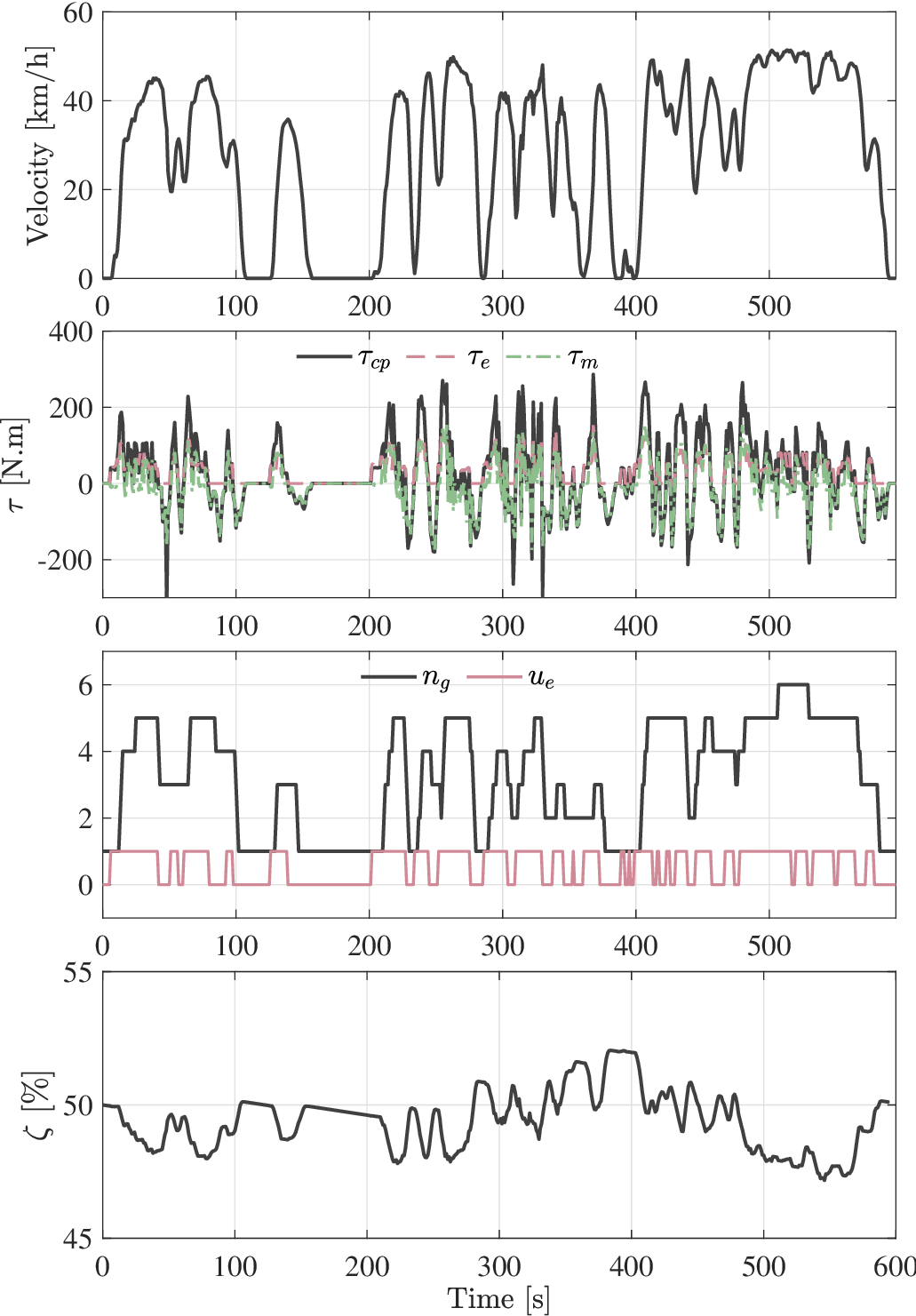} }
\subfigure[][AV cycle]{%
\label{fig:AVD_Cycle_EMS}
\includegraphics[trim= 0cm 0cm 0cm 0cm, clip=true, width=0.4\textwidth]{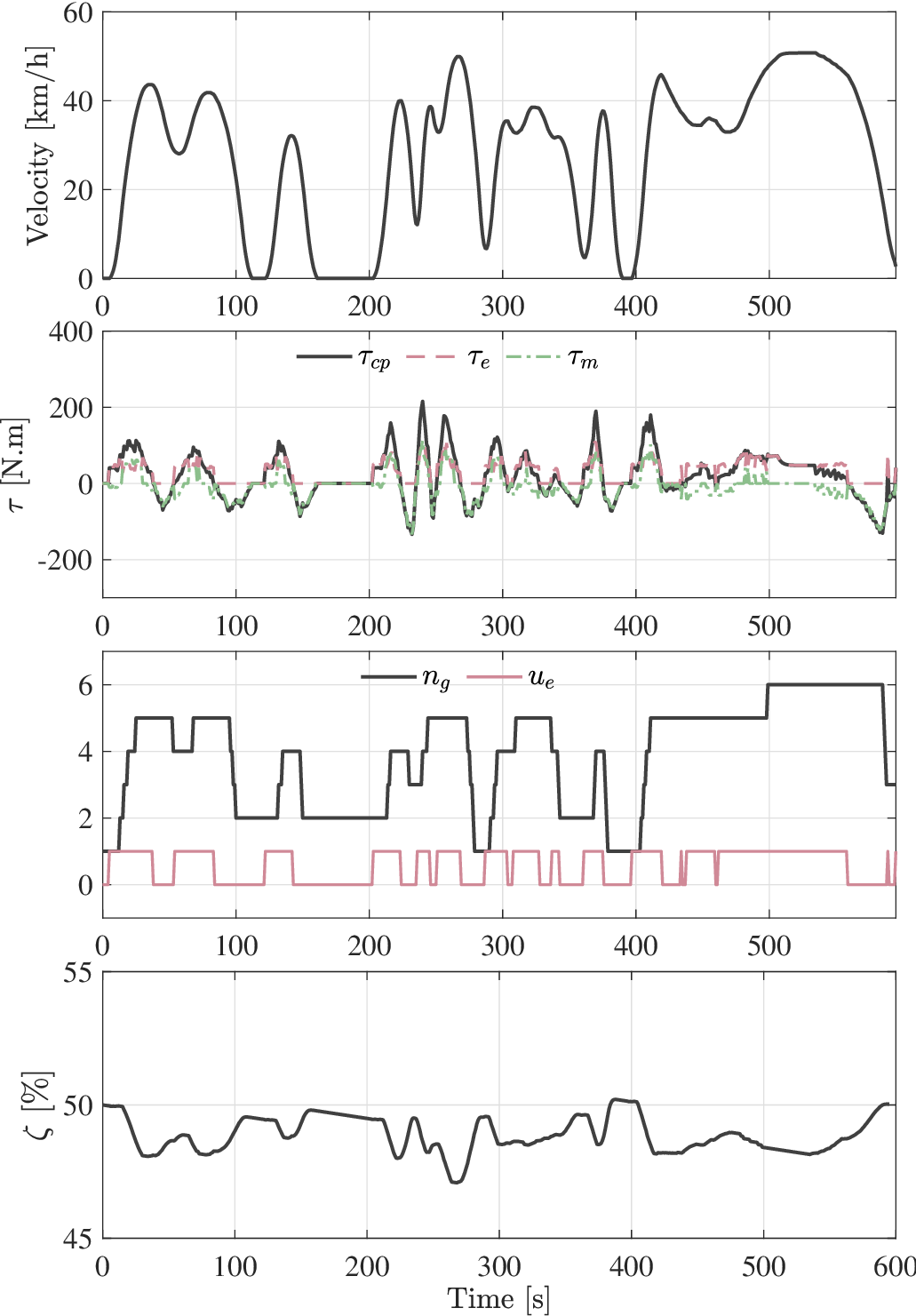} }
\caption{An example of energy management results for HEV with a 30 kW motor for
\subref{fig:HD_Cycle_EMS} a human-driven cycle and 
\subref{fig:AVD_Cycle_EMS} corresponding AV cycle. The top plots are the vehicle velocity, and the second plots show the torque on the torque converter input shaft ($\tau_{cp}$), engine torque ($\tau_e$), and motor torque ($\tau_m$). The third plots represent the gear number ($n_g$) and engine on/off command ($u_e$), and the bottom plots are the battery state of charge.}
\label{fig:EMS_OrigCycle}
\end{figure*}

\begin{align}
\label{eq:cos_func}
\min \Bigg\{ \sum_{k=1}^N \Bigg(\dot{m}_f(k) T_s + \alpha\big|n_g(k)- n_g(k-1)\big| \nonumber \\ +\beta\bigg({\rm max}\big(x_e(k-1),0\big) \bigg)  \Bigg)   \Bigg\}
\end{align}

Subject to constraints
\begin{subequations}
\begin{align}
\label{eq:soc_dyn_optProb}
\zeta(k+1) &= \zeta(k) -\frac{I_b(k)}{3600C_n}T_s\\
\label{eq:gear_dyn_optProb}
n_g(k) &= u_g(k)+n_g(k-1)\\
\label{eq:EngState}
x_e(k) &= u_e(k)\\
\label{eq:SoC_lim_optProb}
\zeta^{\rm min} &\leq \zeta(k) \leq \zeta^{\rm max}\\
\label{eq:Ib_lim_optProb}
I_{b}^{\rm min} &\leq I_{b}(k) \leq I_{b}^{\rm max}\\
\label{eq:Mtr_TA_lim_optProb}
\tau_{m}^{\rm min} &\leq \tau_{m}(k) \leq \tau_{m}^{\rm max}\\
\label{eq:Te_lim_optProb}
\tau_e^{\rm min} &\leq \tau_e(k) \leq \tau_e^{\rm max}\\
\label{eq:We_lim_optProb}
\omega_e^{\rm min} &\leq \omega_e(k) \leq \omega_e^{\rm max}\\
\label{eq:Ug_lim_optProb}
u_g(k) &\in \{-1,0,1\}\\
\label{eq:Ng_lim_optProb}
n_g(k) &\in \{1,2, \hdots, 6\}\\
\label{eq:const_ue}
u_e(k) &\in \{0,1\}\\
\end{align}
\end{subequations}
where $\dot{m}_f$ is the fuel flow rate that is found using the engine fuel consumption map based on engine torque, $\tau_e$, and engine speed, $\omega_e$. $k$ is the problem step time, and $T_s$ is the sample time equal to 1 second. The coefficient $\alpha$ penalizes gear shifts to avoid frequent gear shifts and produce more realistic shift scenarios. The coefficient $\beta$ penalizes the engine's cranking for start-stop. The constraint  (\ref{eq:soc_dyn_optProb}) represents the state of charge dynamics, in which $C_n$ is the battery capacity. The equation (\ref{eq:gear_dyn_optProb}) shows the gear change, and (\ref{eq:EngState}) determines the engine on/off state. The inequality constraints (\ref{eq:SoC_lim_optProb}), (\ref{eq:Ib_lim_optProb}), (\ref{eq:Mtr_TA_lim_optProb}), (\ref{eq:Te_lim_optProb}), (\ref{eq:We_lim_optProb}) respectively curb the battery state of charge, current, motor torque, and engine torque and speed between their minimum and maximum values. 
The constraint (\ref{eq:Ug_lim_optProb}) restricts gear shifts to a single upshift and downshift, (\ref{eq:Ng_lim_optProb}) confines the gear number to 1 to 6, and (\ref{eq:const_ue}) permits turning the engine on and off.

The global optimal energy management problem of (\ref{eq:cos_func}) is solved over all cycles. An example of the results with a 30 kW motor for a human-driven cycle is shown in Fig. \ref{fig:HD_Cycle_EMS}, while the corresponding AV cycle is shown in Fig. \ref{fig:AVD_Cycle_EMS}. The top subplots show the vehicle velocity profile. The second subplots show the torque request on the torque converter input shaft ($\tau_{cp}$, see Fig. \ref{fig_powertrain}), the engine torque, and the motor torque. The third subplots show the gear number and engine on/off command, and the last subplots represent the battery state of charge during the cycle. As expected, the DP-based optimal EMS splits the requested torque between the engine and the motor to minimize the defined performance metric. Furthermore, the engine is turned off when the vehicle is stopped and sometimes when the vehicle is in motion to avoid inefficient engine operation. The next section discusses fuel consumption results in detail.

\section{Discussion}
\label{sec:discussion}
\begin{figure*}[htbp]
\centerline{\includegraphics[trim= 3cm 0cm 0cm 0cm, width=0.95\textwidth]{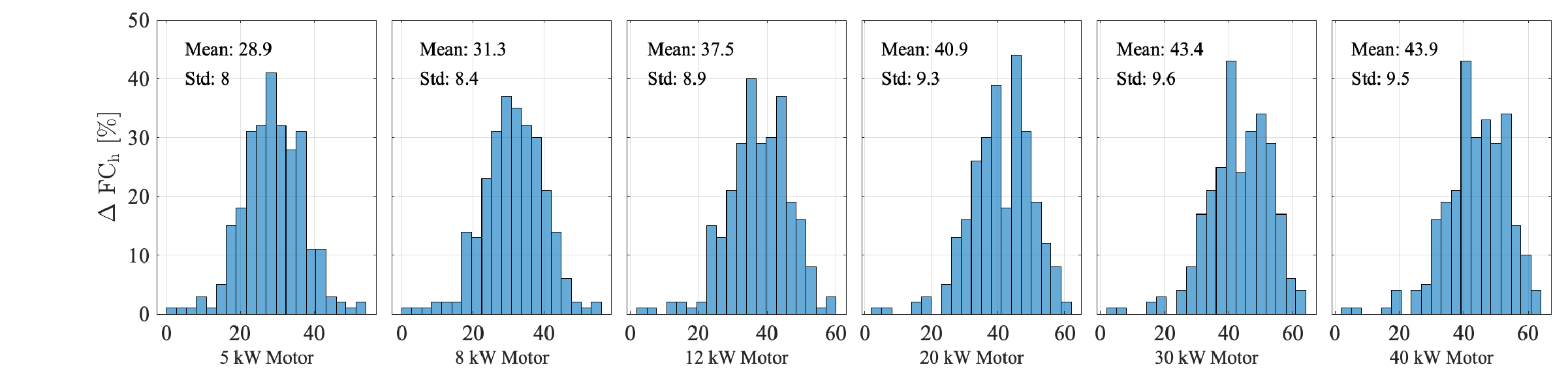}}
\caption{The statistics of fuel consumption reduction due to the powertrain hybridization for different motors sizes for a human driver. The fuel consumption change is computed relative to the baseline ICEV.}
\label{fig:FE_Gain_hyb_HD}
\end{figure*}

\begin{figure*}[htbp]
\centerline{\includegraphics[trim= 3cm 0cm 0cm 0cm, width=0.95\textwidth]{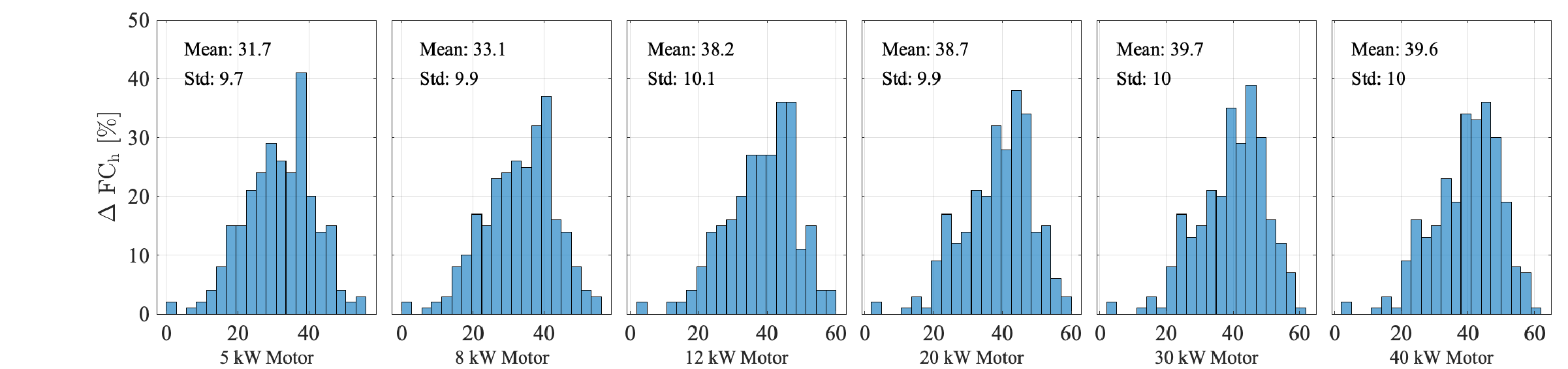}}
\caption{The statistics of fuel consumption reduction due to powertrain hybridization for different motor sizes for autonomous vehicles. The fuel consumption change is computed relative to the baseline ICEV.}
\label{fig:FE_Gain_hyb_AV}
\end{figure*}

\subsection{Fuel Consumption Improvement with Powertrain Hybridization}
The distribution of fuel consumption reduction from powertrain hybridization along with its mean and standard deviation (Std) for different motor sizes is presented in Fig. \ref{fig:FE_Gain_hyb_HD} for a human driver, and  in Fig. \ref{fig:FE_Gain_hyb_AV} for AV cycles. The subplots from the left are respectively for 5 kW, 8 kW, 12 kW, 20 kW, 30 kW, and 40 kW motors. The fuel consumption reduction, $\rm \Delta FC_h$, is computed as
\begin{align}
\rm \Delta FC_h = \frac{FE_{hev}-FE_{ICEV}}{FE_{ICEV}}\times 100
\end{align}
where $\rm FE_{hev}$ is the fuel economy of HEV with the given motor size and $\rm FE_{ICEV}$ is the fuel economy of the baseline ICEV, both in [l/100km]. As seen, hybridization effectively reduces the fuel consumption for both  human drivers and AVs across all motor sizes. However, this reduction is larger for human drivers across all motor sizes. For human drivers, hybridization decreases the fuel consumption by 28.9\%, 31.3\%, 37.5\%, 40.9\%, 43.4\%, and 43.9\% on average with respectively 5 kW, 8 kW, 12 kW, 20 kW, 30 kW, and 40 kW motors, and the corresponding reductions for AV are 31.7\%, 33.1\%, 38.2\%, 38.7\%, 39.7\%, and 39.6\%. 

\begin{figure*}[bth!]%
\centering
\subfigure[][Fuel Consumption Reduction]{%
\label{fig:Driver_FC_Results}
\includegraphics[trim= 0cm 0cm 0cm 0cm, clip=true, width=0.95\textwidth]{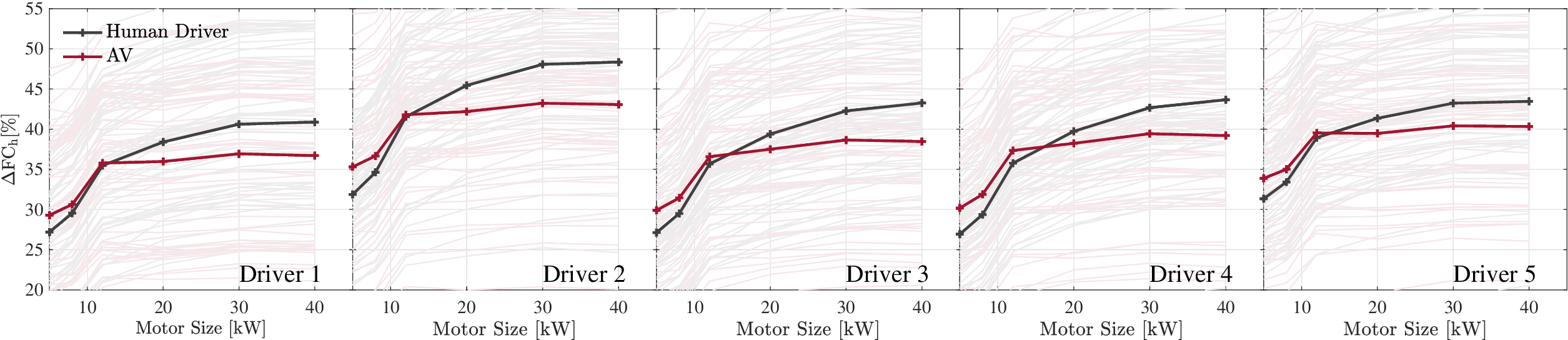} }\\
\subfigure[][Fuel Economy]{%
\label{fig:Driver_FE_Results}
\includegraphics[trim= 0cm 0cm 0cm 0cm, clip=true, width=0.95\textwidth]{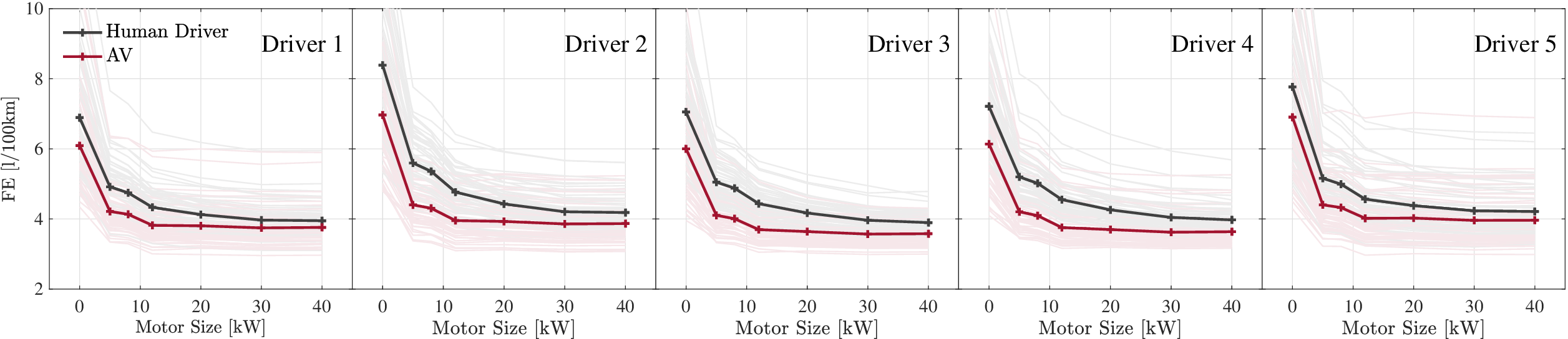} }
\caption{
\subref{fig:Driver_FC_Results} Fuel consumption reduction from the powertrain hybridization
\subref{fig:Driver_FE_Results} and absolute fuel economy numbers for individual drivers. The faded grey and pink lines show drive cycles and the bold colors show the average values.}
\label{fig:FC_D_results}
\end{figure*}

Fig. \ref{fig:Driver_FC_Results} shows $\rm \Delta FC_h$ versus motor size  and Fig. \ref{fig:Driver_FE_Results} shows the fuel economy versus motor size for each driver. The black lines represent the average values for the human driver, and the red lines show the average number for the AV cycles constructed from the same human driver cycles, therefore the same traffic condition. The thin grey and pink lines in the background show the results for individual cycles, respectively for the human driver and AV. These results show that for human drivers, using a larger electric motor results in a higher decrease in fuel consumption for all considered drivers and motor sizes up to 30 kW. For example, using a 30 kW motor instead of an 8 kW motor reduces the FC by around 10\%-14\% on average for different drivers, or upgrading a 12 kW motor to a 30 kW motor reduces the FC by 4\%-7\% on average. In contrast, it is observed that for AV cycles, the FC decrease is small for motor sizes larger than 12 kW, such that increasing the motor size from 12 kW to 40 kW reduces the FC by 1.4\% on average. These results indicate that a 12 kW motor, which can be realized with a 48 V system, is the suitable motor size for an autonomous vehicle in the considered mixed traffic city driving condition because it can produce almost all the fuel economy benefits of a full-size hybrid powertrain. For human-driven cycles, the 30 kW motor seems to be a suitable choice because moving towards a larger motor size, i.e. 40 kW motor, results in an incremental FC decrease of 0.5\%. The underlying reasons for this difference are discussed in detail in Section \ref{sec:Categ_FC}.

\subsection{Fuel Consumption Reduction with Velocity Smoothing}
\label{sec:FC_eco_driving}
\begin{figure}[th!]%
\centering
\includegraphics[trim= 0cm 0cm 0cm 0cm, clip=true,  width=0.45\textwidth]{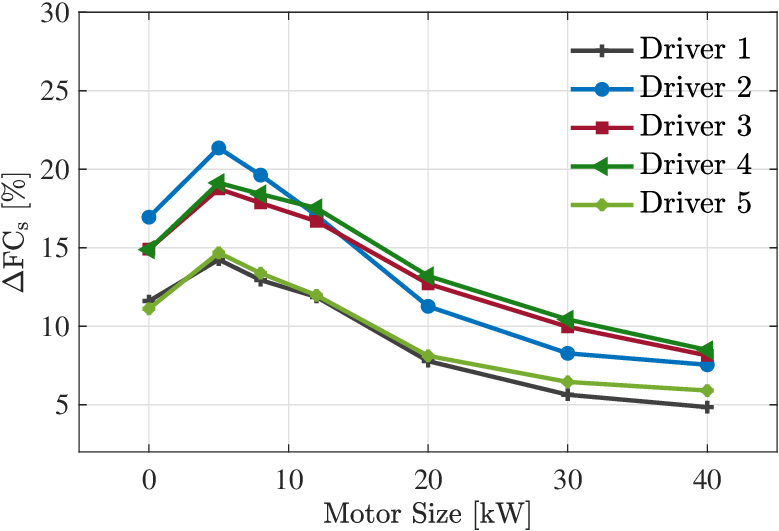} 
\caption{Fuel consumption reduction from velocity smoothing for ICEV and HEVS.}
\label{fig:Driver_Eco_Gain2}
\end{figure}

In this section, we examine the benefits of velocity smoothing enabled with vehicle automation in FC reduction of the baseline ICEV and different HEVs.  Fig. \ref{fig:Driver_Eco_Gain2} shows the average FC gain, $\Delta \rm FC_s$, versus motor size for different drivers, where $\Delta \rm FC_s$ is defined as follows
\begin{align}
\rm \Delta FC_s = \frac{FE_{HD}-FE_{AV}}{FE_{HD}}\times 100
\end{align}
in which $\rm FE_{HD}$ is the average fuel economy of the human driver on all cycles and $\rm FE_{AV}$ is the average fuel economy of AV on the equivalent AV cycles that were realized by optimization (\ref{eq:eco-drive}). The first marker on the left corresponds to the baseline ICEV. 
Velocity smoothing reduces the FC of the baseline ICEV by 11\%-17\% and of the HEVs by 5\%-22\% across all  drivers, however, there is not a monotonous relationship between $\Delta \rm FC_s$ and motor size. As expected, the FC gain for HEVs with large motors is minimum, as hybridization and velocity smoothing work in similar ways to improve the fuel economy of a vehicle. Yet, interestingly, the velocity smoothing benefits are higher for HEVs with small motors compared to the baseline ICEV. The reason is that velocity smoothing reduces the acceleration and deceleration of the vehicle such that a smaller motor size will be capable of recuperating a larger portion of the otherwise wasted vehicle kinetic energy, and therefore increases the effectiveness of these small hybrids. It is also observed that Drivers 2, 3, and 4 that have higher average acceleration and deceleration and therefore,  a more aggressive driving style, benefit more from velocity smoothing across all motor sizes compared to Driver 1 and 5. This observation confirms the fact that an aggressive driving style is detrimental to fuel economy, even with a full hybrid powertrain that has a high capacity for regenerative braking.

\subsection{Categorizing Fuel Consumption Gain}
\label{sec:Categ_FC}

\begin{figure}[bth!]%
\centering
\subfigure[][FC reduction by source]{%
\label{fig:delFC_break}
\includegraphics[trim= 0cm 0cm 0cm 0cm, clip=true, width=0.4\textwidth]{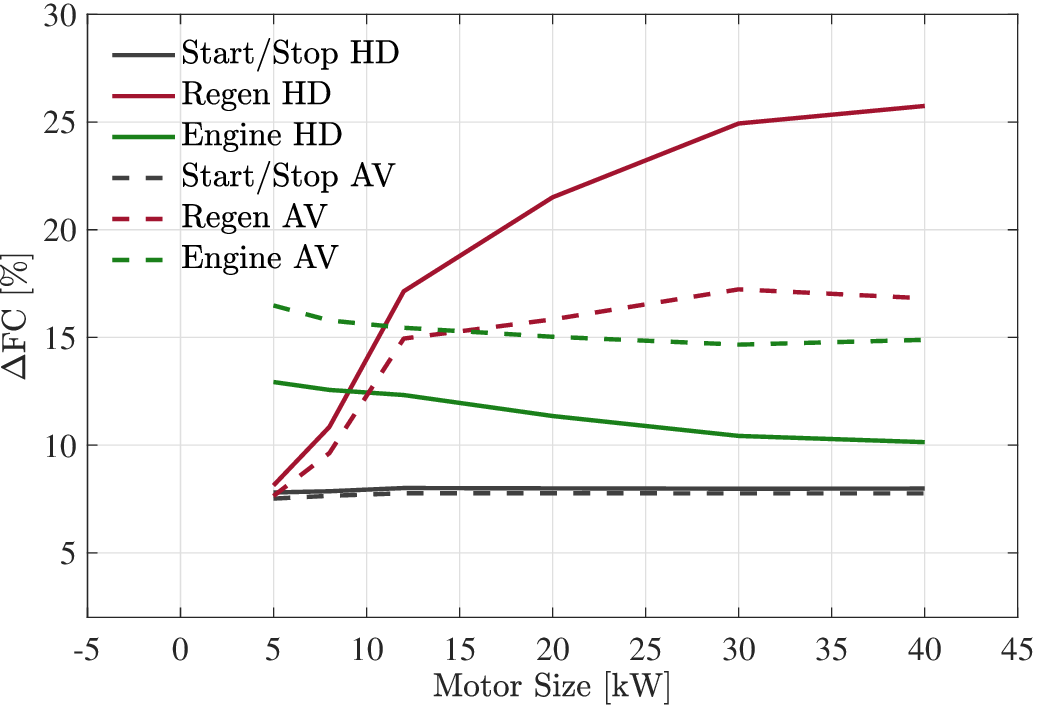} }\\
\subfigure[][Engine Average BSFC]{%
\label{fig:Eng_BSFC_Avg}
\includegraphics[trim= 0cm 0cm 0cm 0cm, clip=true, width=0.4\textwidth]{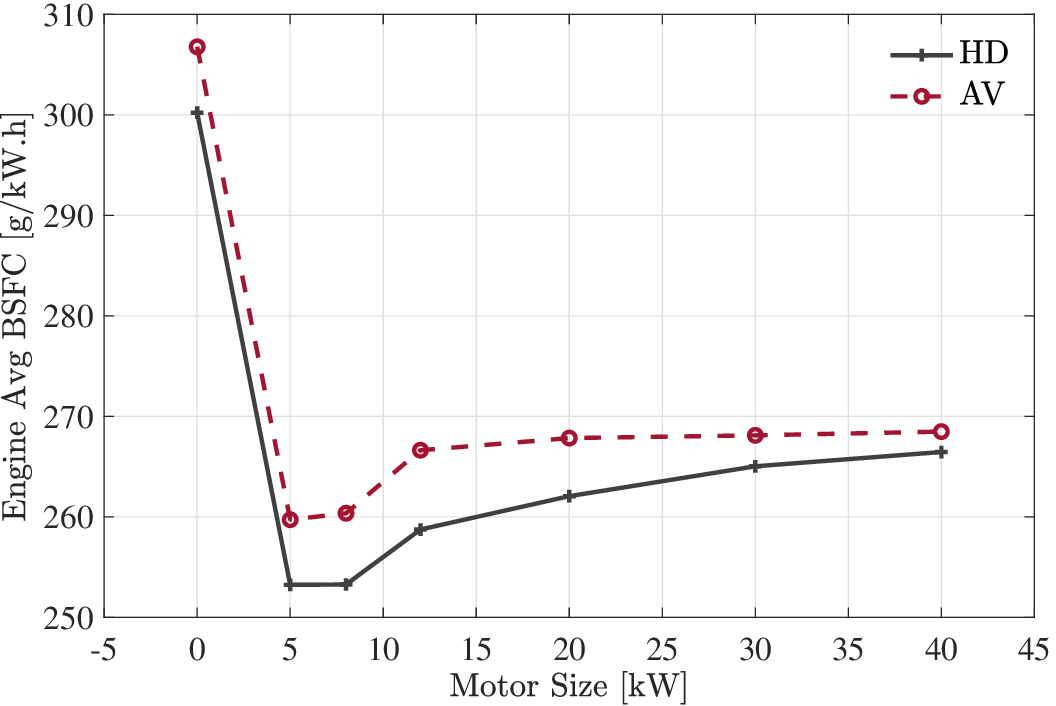} }
\caption{
\subref{fig:delFC_break} Fuel consumption reduction by source, and
\subref{fig:Eng_BSFC_Avg} engine average Brake Specific Fuel Consumption (BSFC). The numbers are the average values over all considered cycles.}
\label{fig:FC_D_results}
\end{figure}

 The results presented in previous parts indicate that even with a small electric motor, the FC reduction is significant on most of the considered cycles. For example, with a human driver more than 25\% FC reduction is achieved on 70\% of the cycles with a 5 kW motor, and on 91\% of the cycles with a 12 kW motor. Similarly, for an AV the same FC reduction is gained on 75\% of the cycles with a 5 kW motor and on 88\% of the cycles with a 12 kW motor. Hybridization reduces the fuel consumption of a vehicle through three main means. First, regenerative braking recovers part of the vehicle's kinetic energy during braking events by operating the motor as a generator and storing the recuperated energy in the battery, which will be used for propulsion or powering auxiliaries later. Second, it turns the engine off when the vehicle is not in motion to avoid extra fuel consumption during idling, and finally, engine operating points are shifted to higher efficiency points by finding the optimum balance between motor torque and engine torque. It is understood that FC reduction in practice will be lower than the numbers generated with DP, mainly due to the sub-optimality of the EMS and the lack of a full driving profile preview. Still, given these results, it is predicted that the FC gain in the studied scenarios will be substantial, because the urban driving cycles studied here, such as the examples shown in Fig. \ref{fig:DC_example1} and \ref{fig:DC_example2}, include many braking and stopping instances, which allow the hybrid powertrain to effectively reduce FC on these cycles.  In this section, the share of each of these mechanisms in FC reduction for both the human driver and the AV is computed. 
 
The recovered energy during regenerative braking events, $\rm E_{rgn}$, is computed as

\begin{align}
\rm  E_{\rm rgn} &= \sum_{k=1}^N\Big(({P}_m(k)+P_e(k))\eta_m(k) T_s\Big)_{P_w(k),P_m(k)<0}
\end{align}
where $P_w$ is the wheel power, ${P}_m$ is the motor power, and $P_e$ is the engine power. This equation  computes the amount of electrical energy that the motor is recuperated during vehicle braking instances ($P_w(k)<0$ and $P_m<0$) from regenerative braking. The reason that the engine power appears in this equation is that simultaneous generation from the engine and regenerative braking is allowed by our EMS formulation. In the next step, the equivalent saved fuel due to regenerative braking is calculated using the battery average round trip efficiency, $\bar{\eta}_b$, the motor motoring average efficiency, $\bar{\eta}_m^m$, and the engine average fuel rate to mechanical power efficiency, $\bar{\eta}_e$ in kJ per gram of burnt fuel, [kJ/g], as
\begin{align}
\rm \Delta FC_{rgn} = \frac{\bar{\eta}_m^m \bar{\eta}_b E_{\rm rgn}}{\bar{\eta}_e}
\end{align}
the values of $\bar{\eta}_b$, $\bar{\eta}_m^m$, $\bar{\eta}_e$ are calculated for each cycle. For example 
\begin{align}
\bar{\eta}_e = \frac{\sum_{k=1}^N \big(P_e(k) T_s)}{\sum_{k=1}^N \big(\dot{m}_f(k) T_s \big) }
\end{align}

The FC reduction due to start-stop, $\rm \Delta FC_{st/sp}$, is computed by integrating the saved fuel due to turning the engine off during vehicle stops, $\rm \sum\Big(\Delta \dot{m}_f(k) T_s\Big)_{v(k),u_e(k) = 0}$, minus the fuel and energy penalty for starting the engine, as follows

\begin{align}
\rm \Delta FC_{st/sp} = \sum\Big(\Delta \dot{m}_f(k) T_s\Big)_{v(k),u_e(k) = 0} -  n_{st/sp} (\beta+ \frac{P_{st/sp}}{\bar{\eta}_m^g \bar{\eta}_b \bar{\eta}_e})
\end{align}
in which  $n_{st/sp}$ is the number of start-stops over a cycle, $\beta$ is the fuel penalty for each start, $P_{st/sp}$ is the electric energy penalty for each vehicle start, and $\bar{\eta}_m^g$ is the average motor efficiency during generation ($P_m < 0$) over the cycle. Note that with our EMS formulation, the engine can turn off when the vehicle is in motion too. In this study, fuel saving during such events is categorized as engine operating point optimization. After computing the fuel saving due to regenerative braking and start-stop, it is assumed that the remaining FC reduction is entirely due to engine operating point optimization.

Fig. \ref{fig:delFC_break} shows the $\Delta \rm FC$ from each source, averaged over all studied cycles. The solid lines show the results for the human drivers and the dashed lines represent AV. The black lines show the FC savings from start/stop, which is around 8\% for human drivers and around 7.8\% for AV. The reason for the slightly lower savings for AV is that the stop time is reduced by a small amount for AV cycles due to the velocity smoothing. The red lines show the FC benefit from regenerative braking, the value of which for a human driver varies from 8.1\% for a 5 kW motor to 25.7\% for a 40 kW motor, and for AV changes from 7.6\% for a 5 kW motor to 16.8\% for a 40 kW motor. The reason for this difference is that with velocity smoothing less energy is available for recuperation during braking events of AV. Finally, the green lines represent the saved fuel due to optimizing the engine operating point. Two trends can be observed here. First, the FC benefit from engine optimization is larger for the AV across all motor sizes. Second, this benefit becomes smaller for larger motor sizes. The average Brake Specific Fuel Consumption (BSFC) of the engine, shown in Fig. \ref{fig:Eng_BSFC_Avg}, confirms these trends. It can be inferred that the velocity smoothing enabled by vehicle automation allows the engine to operate at a higher efficiency, the trend that was also observed in the initial study \cite{nazari2019effectiveness}. The reason for the higher average engine BSFC with a larger motor can be tracked to the sizing problem because the assumed engine will be relatively oversized for HEV powertrains with a large motor. Nevertheless, this effect influences the FC numbers by at most 3\% for human drivers and by 2\% for AV cycles, which are relatively small. On the other hand, using the same engine across all studied powertrains allowed us to simplify and reduce the study variables. The efficiency map of the used engine is published elsewhere \cite{stuhldreher2015downsized} and has a relatively large efficiency sweet spot in medium torque regions. This quality allowed us to use the same engine across different HEVs with minimum FC penalty. Using the total FC gain from start-stop and regenerative braking, as a new measure to determine the suitable motor size for AVs, still, the 12 kW motor is the correct choice since moving from a 12 kW to 40 kW motor only gives a 1.8\% higher fuel saving from these two sources for an AV.

\begin{figure}[htbp]
\centerline{\includegraphics[trim= 0cm 0cm 0cm 0cm, width=0.5\textwidth]{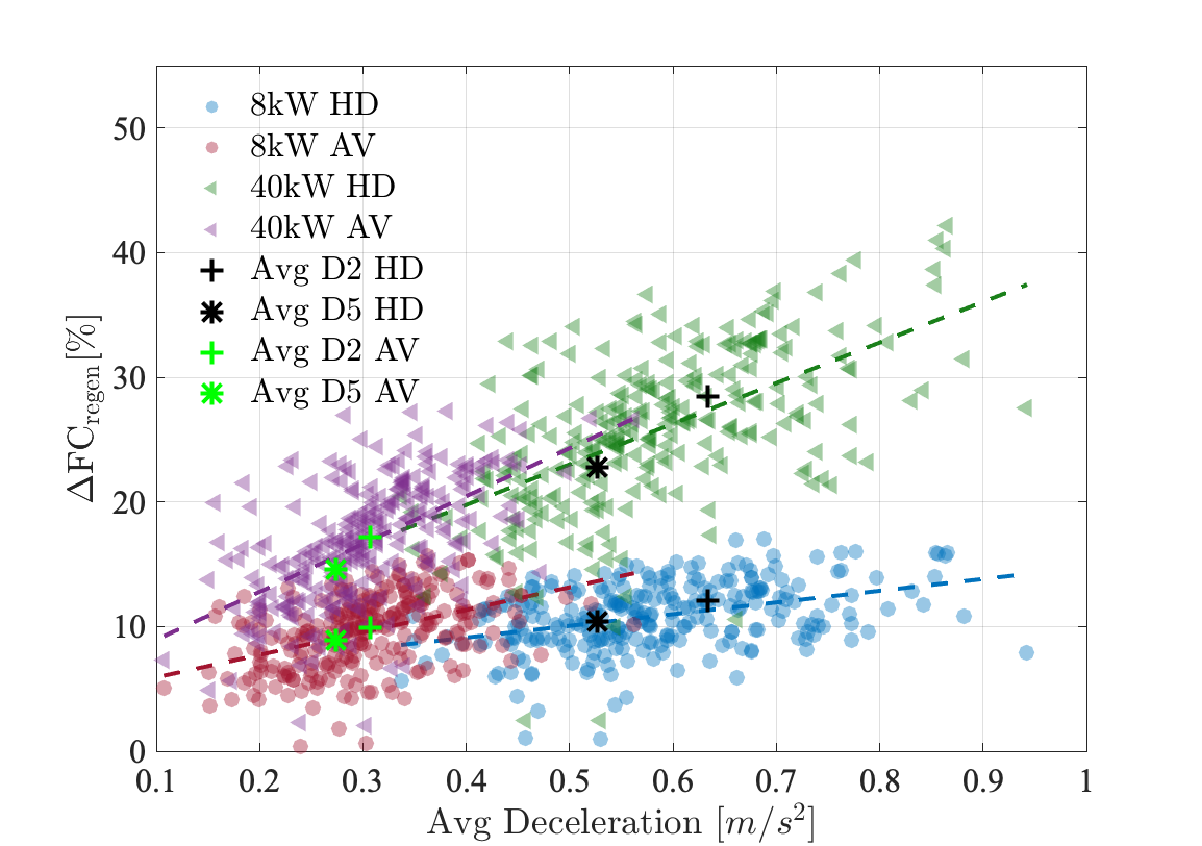}}
\caption{The fuel consumption gain from regenerative braking versus average absolute deceleration of the cycle. Blue circles show HEV with an 8 kW motor and human driver, red circles show HEV with an 8 kW motor and AV, green triangles represent HEV with a 40 kW motor and human driver, and purple triangles indicate HEV with a 40 kW motor and AV. The average numbers for Driver 2 (D2) and Driver 5 (D5) are also shown.}
\label{fig:ACC_FC_Regen}
\end{figure}

The prior results showed that the main mechanism that allows AVs to use a smaller motor compared to human drivers is the lower regenerative braking requirements. In this part, we take a closer look into this synergy. Fig. \ref{fig:ACC_FC_Regen} shows the FC gain from regenerative braking versus cycle average deceleration. The results for the 8 kW motor are shown with circle markers, in blue for human driver, and in red for AV. The numbers for the 40 kW motor are shown with triangle markers, in green for human driver, and in purple for AV. In each case, the dashed line with the same color is the linear fit to the results. The average numbers for Driver 2 (D2), as the most aggressive driver, and Driver 5 (D5), as the least aggressive driver, is shown  respectively with plus and star markers. As seen all lines fitted to the results have positive slope, meaning that a higher fuel saving is possible on cycles with a higher deceleration rate and this relationship becomes more strictly positive with a larger motor. Consequently, as seen Driver 2 which has a larger average deceleration benefits more from the powertrain hybridization. It is also observed that AV cases on average get a higher FC gain compared to the human-driven cycles. This difference is negligible for the 40 kW motor case, but more evident for the 8 kW motor. It can be concluded that the simple velocity smoothing adopted here allows recuperating a larger part of vehicle kinetic energy compared to a cycle with the same average deceleration but with a human driver, and thus velocity smoothing allows us to take advantage of vehicle hybridization even more.

\section{Conclusion}
This work investigated the synergy between velocity optimization, enabled by automated driving, and powertrain hybridization to determine the necessary degree of hybridization for future AVs. We adopted real-world urban driving profiles and computed the corresponding cycles for an AV navigating through the same traffic conditions. The global optimum energy management problem for a parallel hybrid powertrain was solved with different motor sizes and over various human-driven and AV cycles. The results showed that while hybridization will effectively reduce fuel consumption for both AVs and human drivers, AVs need significantly smaller motors to obtain the full benefits of electrification. Although a 30 kW motor was identified as the suitable motor size for human drivers, the simulations showed that an AV needs only a 12 kW motor to get almost the maximum gains. A closer look into the share of different mechanisms in fuel consumption reduction demonstrated that the employed velocity profile optimization reduces the average and distribution of vehicle acceleration/deceleration towards higher effectiveness of regenerative braking, and at the same time operates the engine at a higher average efficiency. These findings indicate that future autonomous vehicles can profit from hybridization with small electric motors that can be realized with low-voltage systems, which are cheap and subject to fewer safety measures compared to high-voltage systems. 

This work also examined the advantages of velocity smoothing for the baseline ICEV and HEVs with different motor sizes and showed that velocity smoothing benefits HEVs with small motor sizes, such as a 5 kW or 8 kW motor more than traditional ICEVs or HEVs with larger motors. Finally, in this study, we aimed at simulating AVs in the same traffic conditions as human drivers, which represents a mixed autonomy situation. Therefore, a future research direction could be to study the optimum motor size in a full autonomy scenario, where there is a larger control over the vehicle velocity and  more smoothing is possible.

\bibliographystyle{IEEEtran}
\bibliography{bibliography}

\appendices
\section{}

\subsection{Vehicle and Driveline Model }
Backward simulation is used here to model the vehicle driveline and powertrain and formulate the global optimal energy management problem. This method determines the vehicle tractive force, $F_t$, using the vehicle speed and effective mass, $v$ and $M_v$ respectively, and the resistance forces on the vehicle, $F_r$, as follows
\begin{subequations}
\begin{align}
F_t &= M_v\frac{dv}{dt} + F_r\\
F_r &= C_0+C_1v+C_2v^2\\
\label{eq:totMass}
M_v &= m_v + \frac{J_w+\gamma^2(J_p)}{r_w^2}
\end{align}
\label{eq:vehMod}
\end{subequations}
in which, $\gamma$ is the gear ratio from the gearbox and final drive ratio, $m_v$ is the vehicle mass, $r_w$ is wheel radius, $J_w$ is the wheel inertia, and $J_p$ is inertial of the powertrain. The coefficients  $C_0$, $C_1$, and $C_2$ are from EPA-reported dynamometer correction factors to calculate the resisting forces on the vehicle, including aerodynamic drag and rolling resistance. In the next step, the wheel rotational speed $\omega_w$, and torque, $\tau_w$,   are determined as
\begin{subequations}
\begin{align}
\omega_w &= \frac{v}{r_w}\\
\tau_w &= F_t r_w
\end{align}
\end{subequations}

Back propagating these values to the gearbox input shaft, the shaft rotational speed, $\omega_g$, and torque, $\tau_g $, are
\begin{subequations}
\begin{align}
\omega_g &= \gamma \omega_w\\
\tau_g &= \frac{\tau_w}{\gamma}\eta_{g}^{-{\rm sgn}(\tau_w)}
\end{align}
\end{subequations}
where $\eta_g$ is the gearbox efficiency.

The position of the clutch between the engine and the motor is used as the control input for engine start/stop, such that it is assumed that when the clutch is engaged, ($u_e=1$), the engine is turned on and when the clutch is open ($u_e=0$), the engine is turned off.  To emulate the torque converter operation in an automatic transmission, the engine speed, $\omega_e$ is limited to operating speeds higher than its idling speed, $\omega_{e, idle}$, when the engine is turned on

\begin{align}
\omega_{e} = \left\{
\begin{array}{lcc}
\max(\omega_g,\omega_{e,idle}) & \text{if} & u_{e}=1\\
0 & \multicolumn{2}{c}{\text{otherwise.}}
\end{array}
\right.
\end{align}

The motor speed is equal to the engine speed when the engine is turned on and is equal to the gearbox inlet shaft speed when the engine is turned off. This assumption is valid by assuming that the low-level controllers are designed to avoid any unnecessary energy loss, such that the torque converter will be locked any time that the engine is turned off,
\begin{align}
\label{eq:Wm_mdl}
\omega_{m} = \left\{
\begin{array}{lcc}
\omega_e & \text{if} & u_{e}=1\\
\omega_g & \multicolumn{2}{c}{\text{otherwise.}}
\end{array}
\right.
\end{align}

The torque converter model computes the turbine torque, $\tau_{\rm tct}$, and the pump torque, $\tau_{\rm tcp}$, using the speed ratio, $SR$, between its output and input shafts,
\begin{subequations}
\begin{align}
SR &= \frac{\omega_{g}}{\omega_{m}}\\
\tau_{\rm tcp} &= \big(\frac{\omega_{\rm tcp}}{K}\big)^2\\
\tau_{\rm tct} &= \tau_{\rm tcp}\times {\rm TR}
\end{align}
\end{subequations}
where $K$ is the K-factor and $\rm TR$ is the torque ratio of the torque converter, both of which are functions of the speed ratio.

The torque on the torque converter input shaft, which is the torque transmitted through both the torque converter clutch and pump, $\tau_{\rm cp}$, is determined as follows
\begin{align}
\label{eq:crnk_Trq}
\tau_{\rm cp} = \left\{
\begin{array}{lcc}
\tau_g & \text{if} & \omega_{e} \leq \omega_g\\
\tau_{\rm tcp}+{\rm max}(0,\tau_{g}-\tau_{\rm tct}) & \multicolumn{2}{c}{\text{otherwise.}}
\end{array}
\right.
\end{align}

This equation imposes the following condition
\begin{itemize}
\item $\tau_{\rm cp}$ is equal to the drive cycle requested torque, when the torque converter is locked, which happens either when the engine clutch is closed ($u_e=1$ and $\omega_e = \omega_g$) or when the engine clutch is open and the engine is turned off ($u_e=0$ and $\omega_e = 0$).
\item When the torque converter clutch is open, meaning that the engine is turned on ($u_e=1$) and $\omega_e > \omega_g$, if the torque from the speed profile is  smaller than the turbine torque ($\tau_g<\tau_{\rm tct}$), the torque converter clutch does not carry any torque and  the extra torque transmitted through the turbine is wasted in the friction brakes. The torque on the input shaft of the torque converter is equal to the pump torque in this situation.
\item Finally, when the torque converter clutch is open and the requested torque is larger than the turbine torque ($\tau_g>\tau_{\rm tct}$), this equation assumes that the remaining needed torque is transferred by the torque converter clutch, which will be in a slipping condition.
\end{itemize}
Note that this work does not model the torque converter clutch position as an independent control input, but rather it is assumed that the torque converter low-level controller is designed to prevent any avoidable energy loss within the system. The model of the torque converter operation is embedded within the equations (\ref{eq:Wm_mdl}) to (\ref{eq:crnk_Trq}).

\subsection{Engine and Motor Model}
The electric motor and the engine  supply the requested torque on the crankshaft. The engine torque is computed as

\begin{align}
\tau_{e} &= {\rm max}( \tau_e^{\rm min}, \tau_{\rm cp} -\tau_{\rm m})\\
\end{align}
in which $\tau_e^{\rm min}$ is the minimum engine torque. This equation implied that during braking events the energy that is not recovered through regenerative braking is wasted in the friction brakes. The maximum allowable engine torque is determined as follows,

\begin{align}
\tau_{e}^{\rm max}  = \left\{
\begin{array}{lcc}
0 & \text{if} & u_{e}=0\\
\tau_{e}^{\rm max}(\omega_e) & \multicolumn{2}{c}{\text{otherwise.}}
\end{array}
\right.
\end{align}

The engine fuel flow rate,  $\dot{m}_f$, is calculated using its Brake Specific Fuel Consumption (BSFC) map and its brake power, $P_e$. A minimum fuel flow rate, $\dot{m}_{f,\rm min}$, of 1.0 $\frac{\rm kg}{\rm h}$ is assumed for the modeled engine
\begin{subequations}
\begin{align}
{\rm BSFC} &= \Gamma(\tau_e, \omega_e)\\
\tilde{\dot{m}}_f &= P_e\times {\rm BSFC}\\
\dot{{m}}_f &= \max (\tilde{\dot{m}}_f, \dot{m}_{f}^{\rm min})
\end{align}
\end{subequations}

The motor electric power, $P_m$ is computed using its electro-mechanical conversion efficiency, $\eta_m$, as follows

\begin{align}
P_m = \tau_m \times \omega_m \times \eta_m^{-\rm{sgn} (\tau_m)}\\
\eta_{m} = \Gamma(\omega_{m}, \tau_{m})
\end{align}

\subsection{Battery Model}
\label{sec:bettery_Model}
An open circuit voltage with a single resistor is used to model the battery state of charge dynamics. The battery power, $P_b$, is computed by adding the motor electric power, the auxiliary power, $P_{\rm aux}$, and the power needed for cranking the engine during each start event, assumed to be 1 $\rm [kJ]$. 
The battery current, $I_b$, is calculated using $P_b$ and the battery voltage, $U_b$,

\begin{subequations}
\begin{align}
P_b &= P_m+P_{\rm aux}+P_{s/s}\times{\rm max}\big(x_e(k-1),0\big)\\
I_b &= \frac{P_b}{U_b(\zeta)}
\end{align}
\end{subequations}

The battery open circuit voltage, $U_{\rm oc}$, which is function of the battery state of charge, $\zeta$, is used to compute the battery terminal voltage as follows 
\begin{subequations}
\begin{align}
U_b &= \frac{1}{2}U_{\rm oc}(\zeta)+\sqrt{\frac{U_{\rm oc}^2(\zeta)}{4}-P_bR_b}\\
\end{align}
\end{subequations}
Finally, a coulomb counting method is used for state of charge dynamics as follows
\begin{align}
\dot{\zeta} = -\frac{I_b}{3600 C_n}
\end{align}
where $C_n$ is the battery capacity in $[A.h]$.

\ifCLASSOPTIONcaptionsoff
  \newpage
\fi

\end{document}